\documentclass[preprint,showpacs,preprintnumbers,amsmath,amssymb]{revtex4}

\usepackage{graphicx}
\usepackage{dcolumn}
\usepackage{bm}

\begin{document}

\preprint{APS}

\title{ Klein tunneling through an oblique barrier in graphene ribbons}
\author{Jung Hyun Oh}
\author{D. Ahn }
\email{dahn@uos.ac.kr}
\affiliation{Institute of Quantum Information Processing and Systems, University of Seoul, Seoul 130-743, Korea}%

\date{\today}

\begin{abstract}
We study a transmission coefficient of graphene nanoribbons with a top gate which acts as an oblique barrier.
Using a Green function method based on the Dirac-like equation,
scattering among transverse modes due to the oblique barrier is taken into account numerically.
In contrast to the 2-dimensional graphene sheet, we find that the pattern of transmission in graphene ribbons depends
strongly on the electronic structure in the region of the barrier.
Consequently, irregular structures in the transmission coefficient
are predicted while perfect transmission is still calculated in the case of metallic graphene
independently of angle and length of the oblique barrier.
\end{abstract}


\maketitle

\section{Introduction}
Recently, there has been considerable interest in transport on a graphene sheet,
a single atomic layer usually pulled out of bulk graphite.\cite{Novoselov,Zhang,Berger}.
Due to its unique two-dimensional closely-packed honeycomb structures,
electrons in it behave like massless Dirac fermions.\cite{Slonczewski,Ando}
One of its fascinating properties is the so-called Klein tunneling where
perfect penetration occurs independently of potential barrier height,
contrast to the conventional, non-relativistic tunneling.\cite{Klein,Dombey,Krekora}
This relativistic effect is basically originated from the gapless electronic dispersion,
which in turn leads to the connection between electron and hole states in graphene.
A sufficiently strong potential, being repulsive for electrons, is attractive for holes and gives rise to
hole states in the barrier to form channels through which electrons can penetrate the barrier.\cite{Dombey}

The idea of the Klein tunneling realized on graphene sheets was
suggested by Katsnelson {\it et al.}\cite{Katsnelson}
and then several experimental attempts have been made to demonstrate the perfect
transmission.\cite{Huard,Gorbachev,Young} 
Up to now, the perfect transmission is addressed 
in terms of sudden phase shift of conductance as a function of magnetic field.\cite{Young}
However, a more direct evidence for Dirac particles may be the incident angle-dependence of transmission coefficient,\cite{Katsnelson} 
which is not realized experimentally yet.
For incident angle $\theta$ and barrier length $D$ the transmission coefficient $T$ of Dirac particles is given by,
\begin{eqnarray}
T=\frac{\cos^2\theta}{1-\cos^2(kD)\sin^2\theta}
\label{tran2D}
\end{eqnarray}
showing oscillating behavior as a function of incident momentum $k$, angle, and barrier length.
However, in the case of a 2-dimensional graphene sheet it is hard to adjust the incident angle $\theta$ because
there are randomly directed particles which result in averaging Eq. (\ref{tran2D}) over the incident angle.
A collimation-gate method to remove the randomness of electrons may be not appropriate because it requires proximate
implementation to a potential barrier less than a mean free path.

A good candidate to resolve the problem is graphene nanoribbons (GNRs)
because electrons in each transverse mode propagate along its axis (i.e., $\theta=0$).
Then, we can adjust the incident angle $\theta$ definitely by adopting an
oblique potential barrier with respect to
the axis of GNRs. 
However, in this case the transverse momentum is quantized
and correspondingly
Dirac particles become massive depending on their occupation to transverse modes.
So, it is interesting to ask about what is the transmission coefficient of GNRs
with an oblique potential barrier and
to see whether the perfect transmission occurs in ribbon structures even under the
oblique potential barrier.

In this work, the Klein tunneling is investigated in graphene nanoribbons (GNRs)
with oblique barriers theoretically.
Based on Green functions taking account scattering among transverse modes
we show that the transmission coefficient also oscillates, however, with rich structures
as a function of incident energy,  angle, and barrier length.
Through the analysis of local density of states, it is found that
rich structures in the transmission coefficients are resulted
from detailed hole states formed in the region of the potential barrier and associated interference.
We also examine effects of inelastic scattering for the experimental realization of the Kleining tunneling
by employing the level broadening in the simplest approximation.

\section{Model and method}
\begin{figure}
\includegraphics[width=0.35\textwidth]{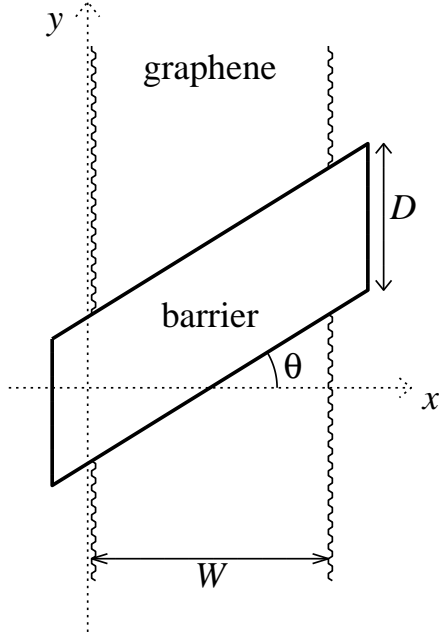}
\caption{A schematic drawing of the device is shown. The oblique gate with an angle $\theta$  is assumed
in a graphene nanoribbon having armchair boundaries. $D$ and $W$ represent length and width of the oblique potential
barrier, respectively.}
\label{system}
\end{figure}

To examine the Klein tunneling we assume a mono-layered graphene ribbon along $y$-direction
with armchair boundaries as shown in Fig. \ref{system}.
It is well known that a graphene ribbon can be metal or 
semiconductor depending on its width $W$.\cite{Brey} We set $W=N a_0/2$ where $2 N$ is a total number
of carbon atoms in a unit cell and $a_0=2.46$\AA ~is the graphene lattice constant.
Then, the system becomes a metal if $N+1$ is a multiple of three, otherwise semiconductors.

The system under consideration is composed of two electrodes and a central device part.
The device part has a finite graphene ribbon with a gate electrode while the lower and upper
electrodes are assumed to be semi-infinite perfect GNRs.
We describe the system with the $4\times4$ Dirac-like equation.
This effective-mass equation is known to give accurate low-energy properties of graphene.\cite{Brey}
With a potential $V(x,y)$ induced by a gate, the Hamiltonian reads as,
\begin{eqnarray}
\left[
\left (
\begin{array}{cc}
-\sigma_x p_x-\sigma_y p_y & 0 \\
0& \sigma_x p_x-\sigma_y p_y   \\
\end{array}
\right)+{\bold 1}V(x,y)\right] 
\psi
 = 
\frac{E}{v_F}
\psi
\label{EoM}
\end{eqnarray}
where $\sigma_{x,y}$ is a Pauli matrix, $v_F\sim 10^6m/sec$ the Fermi velocity, $p_x=-i \hbar\partial_x$
and $p_y=-i \hbar\partial_y$, respectively.
When a gate electrode  is tilted with an angle $\theta$ as shown
in Fig. \ref{system}, the potential barrier is generally a function of $x-$ and $y-$coordinates 
and can be modeled as
$V(x,y)=V(y-(x\!-\!W/2) \tan\theta)$ where
\begin{eqnarray}
V(y) = V_0
\left (
\begin{array}{cc}
1 &{\rm for}~~\mid y\mid \leq \frac{1}{2}(D-d)\\
0 &{\rm for}~~\mid y\mid \geq \frac{1}{2}(D+d)\\
\frac{1}{2}(1-\sin\pi\frac{2\mid y\mid-D}{2d})&{\rm otherwise}
\end{array}
\right..
\end{eqnarray}
Here, we introduce a transition region of the length $d$
to reduce numerical error of a finite difference method used in the followings
as well as for the effective mass equation of Eq. (\ref{EoM}) to be valid.

In the absence of a potential barrier $V(x,y)$,
since the system is a perfect GNR one can solve the equation of motion analytically. 
For instance, the appropriate armchair boundary conditions have been formulated \cite{Brey} and 
used to examine bound states \cite{Trauzettel}.
Wavefunctions are known to be plane waves along both $x$- and $y$ directions as,
\begin{eqnarray}
\psi^0_{n \gamma}(x) = 
\frac{ e^{ik y} }{2\sqrt{W+\frac{a_0}{2}}}
\left (
\begin{array}{c}
  \gamma z_{nk}^\gamma e^{iq_n x} \\
 e^{i q_n x}    \\
-\gamma z_{nk}^\gamma e^{- iq_n x} \\
-e^{-i q_n x}    \\
\end{array}
\right )
\label{psi0}
\end{eqnarray}
where $z_{nk}=\sqrt{q_n-ik}/\sqrt{q_n+ik}$ and  $\gamma=\pm1$ denotes conduction$(+)$ and valence$(-)$ bands of graphene, respectively.
The boundary conditions yield the following quantization for a wave vector in $x$-direction;
\begin{eqnarray}
q_x = q_n = \frac{2\pi}{a_0} \left(\frac{n}{N+1}+\frac{1}{3}\right), ~~~~~n=~{\rm integer}
\label{qx}
\end{eqnarray}
and electronic energy is given by $E=\gamma \hbar v_F \sqrt{q_n^2+k^2}$ for a
propagating wave in $y$-direction with its wave vector $k$.

In general an incident wave with a certain transverse mode
from the lower to upper leads is scattered to other modes
due to a perturbing potential in the device region.
Consequently, one should take into account multiple transverse modes to resolve
a scattering problem.
In our case, we choose the following basis function to describe scattered waves
\begin{eqnarray}
\phi_n^\gamma(x) = 
\frac{1}{2\sqrt{W+\frac{a_0}{2}}}\left (
\begin{array}{c}
\gamma e^{i q_n x} \\
       e^{i q_n x} \\
-\gamma e^{-i q_n x} \\
-e^{-i q_n x} \\
\end{array}
\right )
\label{basis}
\end{eqnarray}
and adopt Green function approach to calculate the transmission coefficient.
Actually, Eq. (\ref{basis}) denotes eigenfunctions of an infinite GNR,
namely Eq. (\ref{psi0}) at $k=0$.
So, since they are orthonormal to each other, i.e.,
$\langle\phi_n^\gamma\mid \phi^{\gamma^\prime}_{n^\prime}\rangle
=\delta_{nn^\prime}\delta_{\gamma\gamma^\prime}$ and
satisfy the armchair boundary conditions,
we can expand the wavefunction of Eq. (\ref{EoM}) as
$\psi(x,y)=\sum_{n\gamma}\chi_{n\gamma}(y) \phi_n^\gamma(x)$.
Resulting equation of motion for $\chi_{n\gamma}(y)$ is summarized as, in a matrix form,
\begin{eqnarray}
\left (E{\bold 1}+{\bold D}(y)+{\bold{\tilde b}}\frac{d}{dy}
\right) {\vec \chi}(y) = 0
\label{EoM1}
\end{eqnarray}
where ${\vec \chi}(y)$ is a column matrix with its component $\chi_{n\gamma}(y)$ and
matrices ${\bold D}$ and ${\bold{\tilde b}}$ are given by,
\begin{eqnarray}
{\bold D}_{n\gamma,n^\prime\gamma^\prime}(y)&=& \hbar v_F \frac{\gamma+\gamma^\prime}{2}
q_n\delta_{nn^\prime}
-\delta_{\gamma\gamma^\prime}
\langle\phi_n^\gamma(x)\mid V(x,y)\mid\phi^\gamma_{n^\prime}(x)\rangle \nonumber\\
{\bold{\tilde b}}_{n\gamma,n^\prime\gamma^\prime}&=&\hbar v_F \frac{\gamma^\prime-\gamma}{2}\delta_{nn^\prime}.
\label{melement}
\end{eqnarray}

Next, we implement a lattice version of Eq. (\ref{EoM1})
by replacing $d/dy$ with a finite difference on a uniform grid.
Detailed results are shown as, especially in a block tridiagonal form,
\begin{eqnarray}
\left (
\begin{array}{ccccccccc}
\cdot &\cdot & \cdot & \cdot & \cdot &\cdot &\cdot &\cdot &\cdot \\
\cdot &  0   &-{\bold b} & {\bold a}_{m-1} & {\bold b} & 0 &\cdot &\cdot &\cdot \\
\cdot &\cdot &    0 &-{\bold b} & {\bold a}_{m} & {\bold b} &  0  &\cdot &\cdot\\
\cdot &\cdot &\cdot  &0 &-{\bold b} & {\bold a}_{m+1} & {\bold b} &  0  &\cdot \\
\cdot &\cdot & \cdot & \cdot & \cdot &\cdot &\cdot &\cdot &\cdot \\
\end{array}
\right )
\left (
\begin{array}{c}
\cdot\\
{\vec \chi}(y_{m-1})\\
{\vec \chi}(y_{m  })\\
{\vec \chi}(y_{m+1})\\
\cdot
\end{array}
\right ) = 0
\label{EoM2}
\end{eqnarray}
where ${\bold a}_m= E{\bold 1}+{\bold D}(y_m)$ and ${\bold b}
={\bold{\tilde b}}/2\Delta$ with a grid spacing $\Delta$.
Following a standard Green function technique,
we define a retarded Green function of the system as inverse of the matrix
in Eq. (\ref{EoM2}) with slightly shifted energy $E\rightarrow E+i\eta$.
Especially, we are interested in the Green function of the device part
which is represented by grid points $m=0,\ldots,M-1$.
For the Green function of the device we truncate the lower- and upper-lead parts
from the original matrix.\cite{Datta} Then effects of the truncation manifest itself
to self-energies as,
\begin{eqnarray}
\left (
\begin{array}{cccccc}
{\bold a}_{0}\!-\!{\bold \Sigma}_L & {\bold b} & 0 &\cdot &\cdot &\cdot \\
-{\bold b} & {\bold a}_{1} & {\bold b} & 0 &\cdot &\cdot \\
\cdot & \cdot & \cdot & \cdot  & \cdot & \cdot \\
\cdot & \cdot & 0& -{\bold b} & {\bold a}_{M-2} & {\bold b} \\
\cdot & \cdot & \cdot & 0 & -{\bold b} &{\bold a}_{M-1}\!-\!{\bold \Sigma}_U  \\
\end{array}
\right )
{\bold G} = {\bold 1}.
\label{GreenD}
\end{eqnarray}
Here, ${\bold \Sigma}^{L,U}$ is called surface self-energy from the lower and upper leads,
respectively, and is related to its Green function,
${\bold G}_{L,U}$ through $\Sigma_{L,U}=-{\bold b}{\bold G}_{L,U}(m_c,m_c){\bold b}$
where $m_c$ is the adjacent index to the device part.
Due to the block tridiagonal form, the Green function for each leads satisfies
the quadratic matrix equation,
$[{\bold a}+{\bold b}{\bold G}_{L,U}(m_c,m_c){\bold b}] {\bold G}_{L,U}(m_c,m_c)={\bold 1}$.
Then, using Eq. (\ref{melement}) we obtain,
\begin{eqnarray}
{\bold \Sigma}^{L,U}_{n\gamma,n^\prime\gamma^\prime}(E) = \frac{1}{2}
\left(1-\sqrt{1-\frac{1}{\Delta^2( E^2/\hbar^2 v_F^2-q_n^2)}}\right)
( E+\gamma\hbar v_F q_n) \delta_{nn^\prime}\delta_{\gamma\gamma^\prime}.
\end{eqnarray}

From the calculated Green function, the local density of states (LDOS) at the index $m$
can be found by
\begin{eqnarray}
{\rm DOS}(m,E) = -\frac{1}{\pi} {\rm Im} {\rm Tr}{\bold G}(m,m)
\end{eqnarray}
where ${\bold G}(m,m)$ is a block matrix of the diagonal Green function at the index $m$.
And the low-field conductance $\sigma$ of the graphene ribbon can be calculated
using the Landauer-B{\" u}ttiker formula\cite{Buttiker}
\begin{eqnarray}
\sigma = \frac{2e^2}{h}\int_{-\infty}^{\infty}
\left (-\frac{\partial f}{\partial E}\right ) T(E) dE
\end{eqnarray}
where $f=(1\!+\!\exp\{(E\!-\!\mu)/k_BT\})^{-1}$ is the Fermi-Dirac distribution function
with the chemical potential $\mu$ and $T(E)$ is the transmission coefficient.
In terms of the most upper-right component of Green function ${\bold G}(0,M\!-\!1)$ 
from Eq. (\ref{GreenD}) the transmission coefficient can be expressed as,
\begin{eqnarray}
T(E) = {\rm Tr}[{\bold G}(0,M\!-\!1)\Gamma^U {\bold G}^\dagger(0,M\!-\!1)\Gamma^L]
\end{eqnarray}
where $\Gamma^{L,U}=i[\Sigma_{L,U}-\Sigma_{L,U}^\dagger]$.

\section{Numerical results}
In this section, we numerically illustrate solutions of the Green function
representing scattered waves from an oblique potential barrier and
related transport properties.
We consider a typical size of the device structure which may be realized experimentally; for instance,
$W=99 a_0$, $D=60a_0$, and $d=30 a_0$.
To include a pattern of an oblique potential barrier, large length of the device part is chosen. 
Actually since the system is in equilibrium, the transmission coefficients are independent
of the total simulation length as long as the scattering potential is described properly in it.
We use 100 transverse modes for the accurate description of scattering waves
and a grid spacing of $\Delta=2$\AA.
So, for the total simulation length of $260a_0$ a dimension of the matrix
Eq. in (\ref{GreenD}) is about $32000$.
Instead of a full-storage scheme
we solve the matrix equation with a standard tridiagonal inversion which is very efficient in
computational demands.

\begin{figure}
\includegraphics[width=0.5\textwidth]{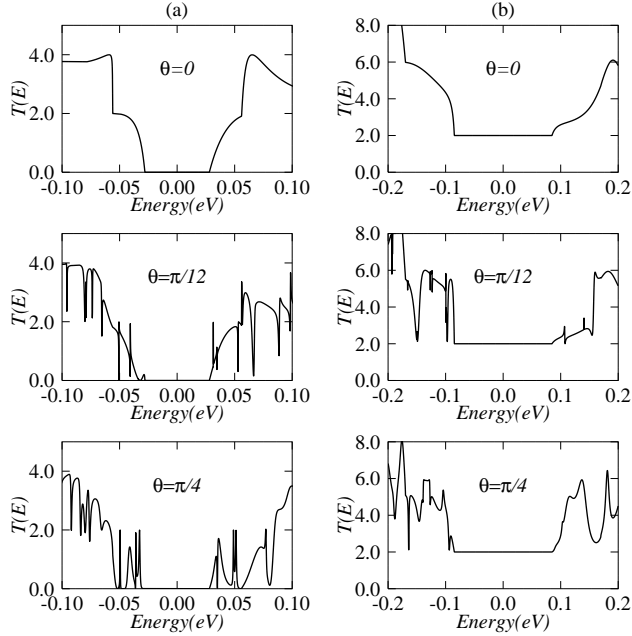}
\caption{Calculated transmission coefficients as a function of incident energy
are compared for (a) semiconducting $(W=99_0)$
and (b) metallic $(W=98a_0)$ cases at oblique angles $\theta=0, \pi/12,$ and $\pi/4$, respectively.
We choose $D=60a_0$, $d=30a_0$, $V_0=0.5$eV for potential barrier parameters with
a grid spacing $\Delta=2$\AA.}
\label{tranm}
\end{figure}
In Fig. \ref{tranm}, we show calculated transmission coefficients through
potential barriers at different oblique angles and compare results
for the cases of metallic and semiconducting graphene ribbons.
For a perpendicular barrier $(\theta=0)$ to incident waves,
one can find the smooth variation of the transmission coefficient as a function of incident energy and
sudden rising at every new occupation to a transverse mode.
In the case of the perfect transmission, the staircase patterns are expected and thus
the deviated ones in the figures imply back-scattered electrons due to the potential barrier.
It is found that the perfect transmission occurs only in the lowest subband in the metallic case.
Actually, this subband has $q_n=0$ and
thereby has the linear dispersion like that in a 2-dimensional case,
which results in the perfect penetration as implied in Eq. (\ref{tran2D}).

As the oblique angle $\theta$ increases, rich structures are found 
as shown in the second and third rows of Fig. \ref{tranm}.
Comparing results at $\theta=0$, one can see that the transmission 
coefficient is very sensitive to the incident energy; many peaks and dips
appear in the small range of energy.
 For $\theta=\pi/4$ cases, calculation results are even similar to resonant tunneling in usual tunneling problems.
On the other hand, it is very interesting to observe robust behavior of the lowest subband for the metallic case
against the obliqueness of the potential barrier.
The perfect transmission is retained within the presence of a single mode, i.e., below the onset of the second transverse mode
regardless of the oblique angle.

This behavior can be understood when we consider the scattering problem in terms of the Fermi golden rule,
the first order perturbation based on eigenstates of Eq. (\ref{psi0}) for infinite GNRs.
If the incident energy is small to occupy only the lowest mode, electrons cannot be scattered by the oblique barrier
to higher modes due to the energy conservation.
This in turn means that for an incident momentum $k$
electrons can be scattered forwardly $(k)$ or backwardly $(-k)$ within the same mode.
However, in the case of a metallic graphene ribbon the lowest mode has a momentum of $q_n=0$ and thus
$z_{kn}$ in Eq. (\ref{psi0}) has  exactly the out-of phase for the backward-scattering relative to the incident one.
Consequently, there is no the backward scattering in the metallic GNRs independently of the oblique angle.
This is not the case for semiconducting GNRs because the lowest mode has $q_n\neq 0$ and the out-of phase cannot be achieved
for backward-scattered waves.

\begin{figure}
\includegraphics[width=0.45\textwidth]{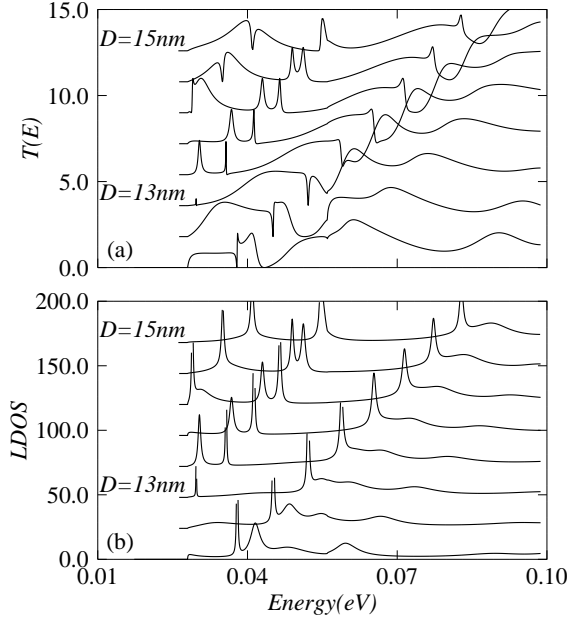}
\caption{
For semiconducting GNR $(W=99_0)$ with the oblique potential barrier $(\theta=\pi/4)$,
we plot calculated transmission coefficients in $(a)$ and local density of states in $(b)$, respectively,
for various barrier lengths.
All curves are vertically offset for clarity and other parameters are the same as those in Fig. \ref{tranm}.}
\label{wdep}
\end{figure}
The barrier-length dependence of the transmission coefficients is examined in Fig. \ref{wdep}-(a).
Contrast to oscillating behavior of the 2-dimensional graphene sheet as in Eq. (\ref{tran2D}),
more peaks and dips appear in the transmission coefficient and
are found to be blue-shifted as we increase the barrier length.
For understanding of the calculated results we first note the blue shifted behavior.
Interference effects are not appropriate to explain it because wavelengths of incident waves usually obey
the geometrical relation of $k D=\sqrt{E^2-q_n^2}D = constant$,
which implies the red-shifted patterns with increase of the barrier length.
So, as an usual explanation of the Klein tunneling
we think about the connection to hole states in the barrier region.
For this we examine the local density of states at the midst of the barrier
and show calculated results in Fig. \ref{wdep}-(b).
Interestingly one can find that the variation of the local density of states
is similar to that of the transmission as a whole.
That is, peaks representing localized states become blue-shifted
and more peaks at lower energy side appear as the barrier length increases.
This is in accordance with usual behavior of states in a quantum well as its size is varied.
Furthermore one can see that the interval between peaks is approximately
equal to the energy-level difference in a quantum well,
i.e., $\Delta E=\hbar v_F /{\rm max}\{D,W\}$ inferred from Eq. (\ref{qx}).
Consequently, we attribute the variation of the transmission coefficient to that of hole state in the barrier region and
peaks in the density of states contribute largely to the transmission coefficient.

\begin{figure}
\includegraphics[width=0.45\textwidth]{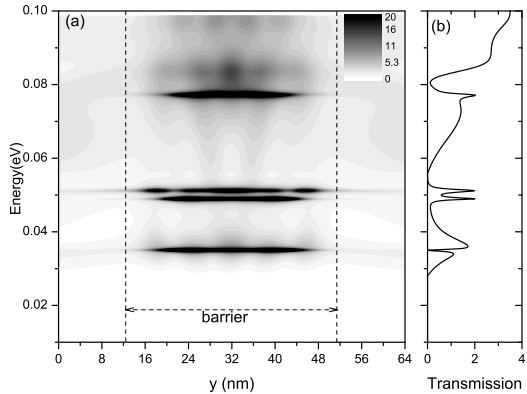}
\caption{
For semiconducting GNR $(W=99a_0)$ with the oblique potential barrier $(\theta=\pi/4)$,
gray-scaled local density of states are depicted as a function of energy along the GNR axis.
$D=60a_0$, $d=30a_0$, and $V_0=0.5$eV are chosen.}
\label{fulldos}
\end{figure}

However, it is noted that all peaks in the density of states are not reflected in the transmission.
In Fig. \ref{fulldos}, we show the local density of states as a function of energy along the device axis.
One can see that high density of states in the barrier region causes a peak in the transmission coefficient.
On the other hand, the dense region in the local density of states, for instance around $E=0.03$eV,
corresponds to a dip in the transmission, which means that detailed transmission is also affected by interference;
We attribute this to destructive interference.

Now, we examine effects of inelastic scattering on the transmission occurring in graphene ribbons possibly from
phonon, edge roughness, and impurities.
To estimate its effects roughly, we adopt the simplest approximation where a diagonal self-energy
representing scattering is added to the Hamiltonian independently of position and energy.
This effect is equivalent to the level broadening by setting a finite value $\eta$ in the retarded Green function.
The value $\eta$ is related to the scattering rate $\gamma$ via the relation of
$\eta = \hbar\gamma$.\cite{Lake,Jin}
According to Ref. 19, the scattering rates $\gamma$ are calculated to have values
ranging about $1\sim 100$THz, equal to $\eta = 0.66\sim 66$meV at $T=300$K
for $W=5$nm or smaller values for wider ribbons and lower temperature.

\begin{figure}
\includegraphics[width=0.45\textwidth]{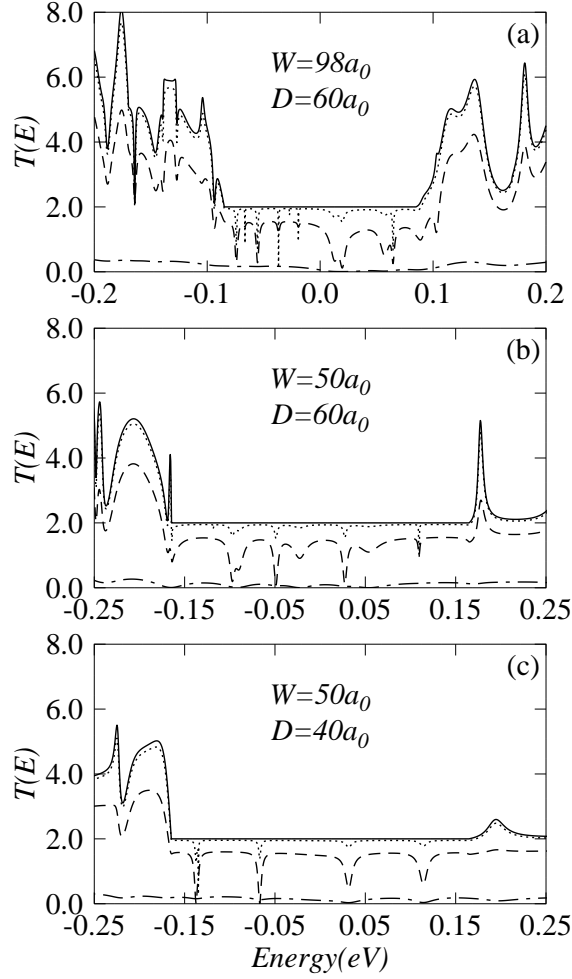}
\caption{ Effects of the level broadening are examined for metallic GNRs with an oblique barrier
$\theta=\pi/4$ for $\eta=0$ (solid line), $0.1$meV (dotted), $1$meV (dashed), and $10$meV (dot-dahsed), respectively.
We use different sizes of the barrier in each panel while other calculation parameters are the same as those
in Fig. \ref{tranm}.
}
\label{brd}
\end{figure}
In Fig. \ref{brd} we show effects of the level broadening on the transmission coefficients
for metallic graphene ribbons at $\theta=\pi/4$.
It is found that the pattern of transmission becomes smeared 
with increasing the broadening together with accompanying many new dips in the region of the perfect transmission,
and eventually has no structures up to $\eta=10$meV.
We also find  similar behavior for different size of GNRs as shown in the second and third rows of Fig. \ref{tranm},
implying that the new dips are irrelevant of the energy difference between transverse modes.
On the other hand, in the case of $\theta=0$, no new peaks are developed while the transmission coefficient are still suppressed
with increase of the broadening (not shown in the figure).  Through the analysis of the local density of states,
we see that positions of the dips correspond to those of abundant density of states in the barrier region.

According to the Fermi golden rule, the level broadening gives rise to scattering to more diverse states
by releasing the energy conservation from the delta function to the Lorentzian one.
So, the scattering matrix is averaged over the energy interval of $\Delta {\cal E} \sim {\rm min}\{\eta,\hbar v_F/D,\hbar v_F/W \tan\theta\}$.
For this reason, the out-of phase of the back-scattered waves is no longer achieved and
thereby the transmission coefficient becomes suppressed.
However, we find that the newly developed dips reflecting more strong suppression cannot be explained by the Fermi golden rule
and may be attributed to higher-order perturbations representing multiple scattering and associated interference.
As a results, since the pattern of transmission is washed out for the level broadening about $\eta=10$meV,
it is necessary to reduce scattering rates for the experimental realization of Dirac particles,
for instance, by lowering temperature and preparing clean GNRs.

\section{Summary}

In this work, we studied the transmission coefficient of graphene ribbons
with an oblique barrier based on the Dirac-like equation.
In contrast to the 2-dimensional graphene sheet,
the transmission in graphene ribbons is found to depend strongly on the electronic structure
in the region of barriers. Consequently, irregular structures in the transmission coefficient
are predicted, however the perfect transmission is shown in the case of metallic graphene
independently of angle, width, length of oblique barriers.
This behavior still demonstrates the nature of Dirac particles in graphene ribbons because
very large potential barrier is assumed.
We also examine effects of scattering by employing the simplest approximation. 

\acknowledgments{This work was supported by the University of Seoul and Seoul Metropolitan government}

\end{document}